\documentclass[12pt]{article}

\usepackage{amsmath}
\usepackage{amssymb}
\usepackage{graphicx}
\usepackage{epstopdf}
\usepackage{epsfig}
\usepackage{epsf}
\usepackage{rotating}
\usepackage{float}

\usepackage{caption}
\usepackage{subcaption}
\usepackage{lineno}
\newcommand\pubnumber{DPF2013-97}
\newcommand\pubdate{\today}
\def\napoli{$^{\textit{a}}$\textit{ Department of Physics and Astronomy,}\\
\textit{ University of New Mexico, Albuquerque, NM 87131, USA\support}\\
$^{\textit{b}}$\textit{Arizona State University, Tempe, AZ 85287, USA}}
\def\support{\footnote{Work supported by the U.S. Department of Energy.}}
\def\Title#1{\begin{center} {\Large #1 } \end{center}}
\def\Author#1{\begin{center}{ \sc #1} \end{center}}
\def\Address#1{\begin{center}{ \it #1} \end{center}}

\newcommand\pubblock{\rightline{\begin{tabular}{l} \pubnumber\\
         \pubdate  \end{tabular}}}
\newenvironment{Abstract}{\begin{quotation}  }{\end{quotation}}
\newenvironment{Presented}{\begin{quotation} \begin{center} 
             PRESENTED AT\end{center}\bigskip 
      \begin{center}\begin{large}}{\end{large}\end{center} \end{quotation}}

\def\beq{\begin{equation}}
\def\eeq#1{\label{#1}\end{equation}}
\def\eeqn{\end{equation}}

\def\beqa{\begin{eqnarray}}
\def\eeqa#1{\label{#1}\end{eqnarray}}
\def\eeqan{\end{eqnarray}}

\let\bar=\overbar

\def\Dslash{\not{\hbox{\kern-4pt $D$}}}
\def\dslash{\not{\hbox{\kern-2pt $\del$}}}

\def\msb{{\bar{\ssstyle M \kern -1pt S}}}

\begin{document}
\begin{titlepage}
\pubblock

\vfill
\Title{A Method for Real Time Monitoring of Charged Particle Beam Profile and Fluence}
\vfill
\Author{P.~Palni$^{\textit{a}}$, M.~Hoeferkamp$^{\textit{a}}$, A.~Taylor$^{\textit{a}}$, S. Vora$^{\textit{b}}$,\\ H.~McDuff$^{\textit{a}}$, Q.~Gu$^{\textit{a}}$, and S.~Seidel$^{\textit{a}}$}
\Address{\napoli}
\vfill
\date{}                                           % Activate to display a given date or no date

%\maketitle
%
\begin{Abstract}
Detectors planned for use at the Large Hadron Collider will operate in a radiation field produced by beam collisions.  To predict the radiation damage to the components of the detectors, prototype devices are irradiated at test beam facilities that reproduce the radiation conditions expected.  The profile of the test beam and the fluence applied per unit time must be known. Techniques such as thin metal foil activation and radiographic image analysis have been used to measure these; however, some of these techniques do not operate in real time, have low sensitivity, or have large uncertainties. We have developed a technique to monitor in real time the beam profile and fluence using an array of $p-i-n$ semiconductor diodes whose forward voltage is linear with fluence over the fluence regime relevant to, for example, tracking in the LHC Upgrade era. We have demonstrated this technique in the 800 MeV proton beam at the LANSCE facility of Los Alamos National Laboratory.

\end{Abstract}
\vfill
\begin{Presented}
DPF 2013\\
The Meeting of the American Physical Society\\
Division of Particles and Fields\\
Santa Cruz, California, August 13--17, 2013\\
\end{Presented}
\vfill
\end{titlepage}

\def\thefootnote{\fnsymbol{footnote}}
\setcounter{footnote}{0}

\section{Introduction}
Development of instrumentation often requires study of the interaction between high energy charged particles and materials. The energy transferred by charged beams through ionization and lattice displacement can lead to a loss of performance and accelerated aging of structural materials and electronic devices. Devices for the LHC or another future collider are typically tested for this sort of effect by being placed in a charged beam. We have developed a technique for real time measurement of the beam profile and fluence. This is an alternative to other methods such as thin metal foil activation [1], radiographic image analysis [2], flying wire  [3], and Faraday cups [4], some of which are either not read concurrently with the beam operation, have larger uncertainties, or have lower sensitivity.

\section{Description of the Diode Array}

We construct an array of OSRAM BPW34F $p-i-n$ diodes [5] to characterize the charged particle beam. When $p-i-n$ diodes with bases manufactured from high resistivity $n$-type silicon are operated under the conditions of low injection, the concentration of carriers in the base region varies such that the resistivity $\rho$ varies as a function of charged particle fluence $\Phi$, as  $\rho= \rho_{0} e^{\Phi /K_{\rho}}$. Here $\rho_{0}$ is the initial equilibrium resistivity of silicon before irradiation and the coefficient $K_{\rho}$ has a value between 400 and 3000 cm$^{-2}$ for different silicon materials [6].\\
\\
The forward voltage across the diode increases linearly with the fluence when supplied with a constant forward current. The diode's forward voltage response at 1 mA, as a function of fluence, is shown in Figure 1 for exposure to 23 GeV protons and 0.8 MeV neutrons. On this graph, the response of the $p-i-n$ diodes to the proton damage is linear in the fluence range from $2 \times 10^{12}$ to $10^{15}$ 1 MeV neutron equivalent (neq) per cm$^2$ before reaching saturation [7].  In the fluence region below 2 $\times~10^{12}$ neq per cm$^2$ (not studied here), high-sensitivity diodes from CMRP would provide a similar linear characteristic [8]. Advantages of using an array of $p-i-n$ diodes to measure the fluence include ease of readout, high spatial resolution, wide range of fluence response, independence of device orientation, dose-rate independence, and commercial availability at very low cost. A disadvantage of the diode is its temperature dependence. We minimize this disadvantage by sourcing the 1 mA current needed to operate them in short (130 ms) pulses.

 \begin{figure}[t!]
\begin{center}
\includegraphics[height=3in, width=4in]{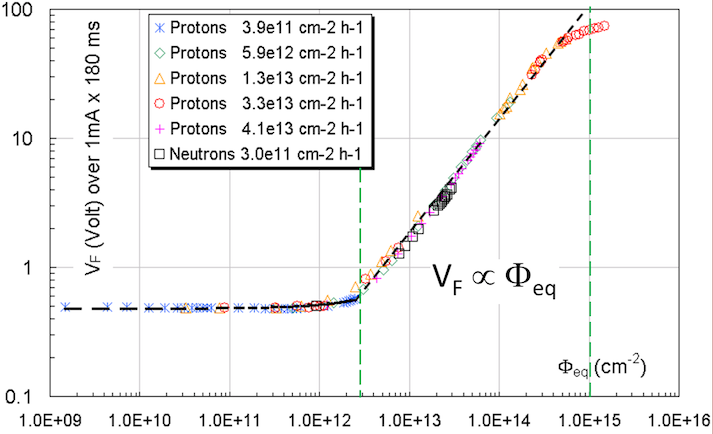}
\caption{ Forward voltage of a single OSRAM diode as a function of fluence in 1 MeV neutron equivalent cm$^{-2}$ [7].}
%\label{ }
\end{center}
\end{figure}

\section{Diode Array Readout Hardware and Software}

\begin{figure}
\centering
  \includegraphics[height=3in,width=3in]{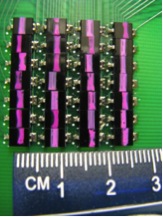}
  \caption{The front side of the diode array.}
 
\end{figure}

\begin{figure}[t!]
\centering
  \includegraphics[height=4.3in,width=6in]{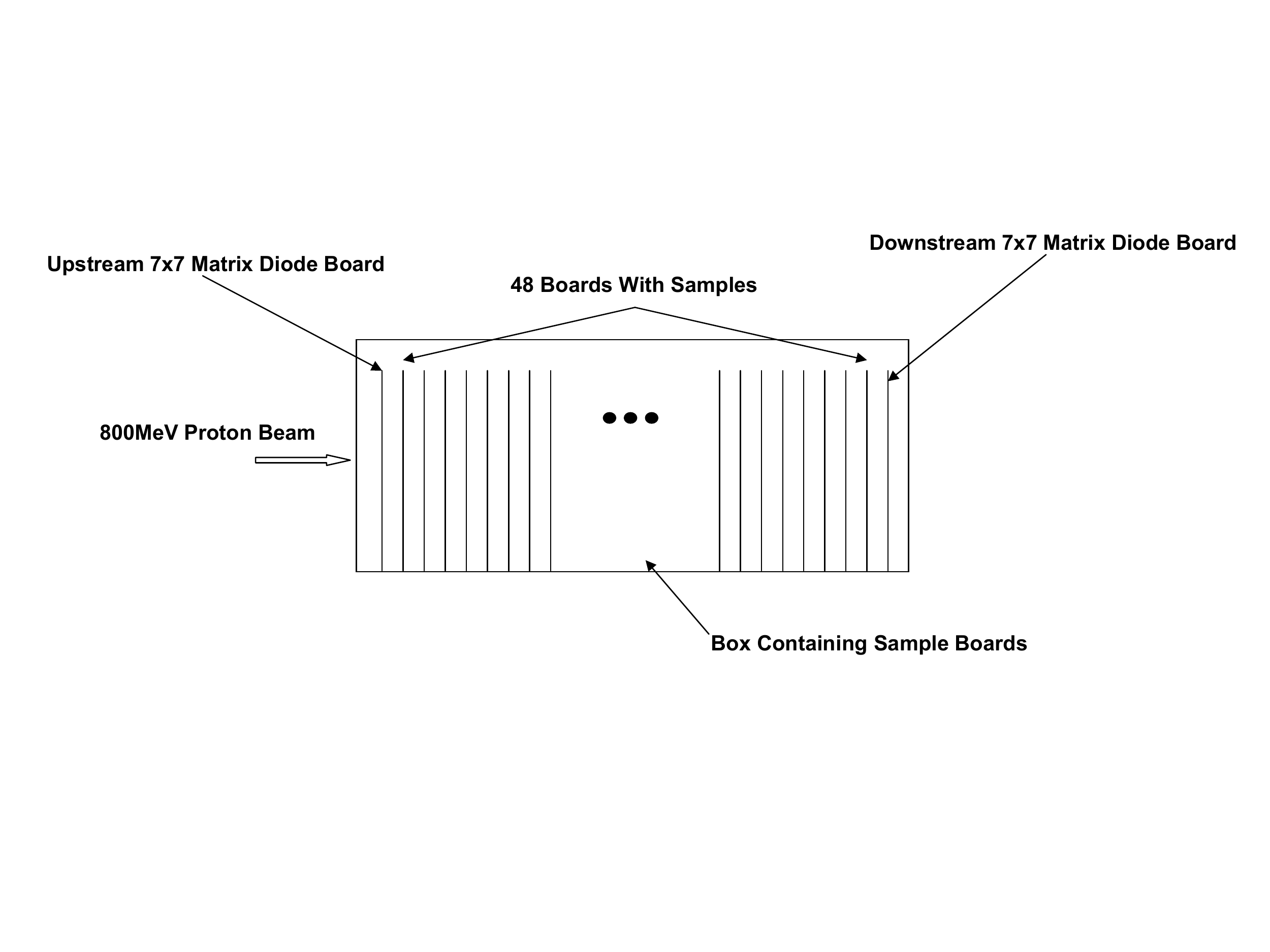}
  \caption{Setup layout of a stack box.}
  \label{fig:sub1}
\end{figure}
The diodes are soldered to back-to-back metalized pads on the two sides of a G10 board.  Four columns of seven diodes each are on one side, and three columns of seven diodes each are interleaved between them on the other side, producing a 7x7 array with nearly complete coverage of a 2.5 cm$^{2}$ region when operated altogether (see Figure 2). The active area of each BPW34F diode is 2.65 mm x 2.65 mm, and the pitch between their centers is 3.8 mm. The board can be placed in a stack box (see Figure 3) with the devices under test (DUT). Custom automated diode scanner software using LabVIEW is capable of scanning 49 channels quickly and remotely without stoppage of the beam. To scan a specific channel, a source measure unit sources a pulse of current and reads out the forward voltage across the $p-n$ junction (see Figure 4). No special environment is required for these measurements.\\
\\
\begin{figure}[t!]
\centering

  \includegraphics[height=3in,width=3.6in]{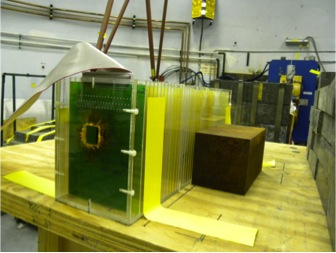}
  \caption{Stack box along the proton beam. The diode array is attached to cables in the first position.}
%\caption{ Experimental setup of the stack box at Target 2 at LANSCE.}
\label{fig:test}
\end{figure}
Our diode array system uses a Keithley 2410 SourceMeter, a Keithley 706 Scanner, and a LabVIEW application. The LabVIEW code controls the setup and functioning of the SourceMeter and Scanner. In general the SourceMeter is set to source a 1 mA constant current while measuring the forward voltage of the selected diode. The Scanner selects each of the diodes as it is pulsed and reads them out one at a time. The total time per diode measurement is approximately 130 ms.

\section{Calibration and Example Implementation}
\begin{figure}
\centering
  \includegraphics[height=3in,width=3in]{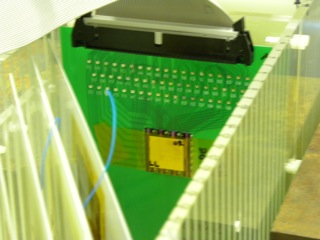}
  \caption{ Aluminum foil matrix attached to the diode array.}
\end{figure}

Two diode arrays were irradiated at the Los Alamos Neutron Science Center (LANSCE) in September 2012. The accelerator provides bunches of $5\times10^{11}$ protons per macro-pulse at an energy of 800 MeV. The diameter of the proton beam spot is about 2 cm. This proton beam is maintained at a constant current of $80~\mu$A. A useful configuration is to place one array at each end of the stack to monitor beam depletion. Figure 5 shows the DUT stack box in the beam including one of the diode arrays. The electrical connections used for the beam profile measurement are shown in Figure 6. The arrays were read out over a 30 m cable after fluences of about 4 $ \times 10^{13}$, 2 $ \times 10^{14}$, 3.2 $\times 10^{14}$, and 8.2 $ \times 10^{14}$ neq per cm$^2$.\\
\\
We used aluminum foil activation to calibrate the diode response to fluence from the diode array for the LANSCE 800 MeV proton beam. A foil of size 2x2 cm$^{2}$ was attached directly to the diode array as shown in Figure 5. We then measured the activity of its central 1 cm$^{2}$ region and converted this to the proton fluence received by it. We also used four aluminum foils adjacent to the diode array in the stack box.  At various points in the irradiation, the diode array was read out and one of the foils was removed at the same time. Figure 7 shows our measurements of fluence (from foil activation) and voltage (from the adjacent diodes). From the fitted line, we obtain a linearity coefficient $c= (6.786 \pm 1.090) \times 10^{-14}$ V/cm$^{-2}$ for 800 MeV protons, which converts to $(9.558 \pm 1.536) \times 10^{-14}$ V/cm$^{-2}$ for neq, using the hardness factor $k= 0.71$. This is consistent with the value of $(10.989  \pm 2.197) \times 10^{-14}$ V/cm$^{-2}$ obtained in [9].\\
\\
As a test of the effectiveness of this technique, the diode array measurements at the upstream and downstream ends of a stack of approximately forty 300 $\mu$m thick silicon sensors were compared four times over the period during which 3.7 x 10$^{15}$ 800 MeV protons were delivered. The results of this study are shown in Figure 8, where one sees the beam profile spread as a result of scatters within the stack.
The full width at half maximum (FWHM) measured by the upstream and downstream diode arrays is shown in Table 1 separately for the X and Y dimensions.

\begin {table}[H]
\begin{center}

\begin{tabular}{|c|c|c|c|c|}
\hline Proton & Upstream array & Upstream array & Downstream array & Downstream array \\ 
 Pulses & FWHM X (mm) & FWHM Y (mm) & FWHM X (mm) & FWHM Y (mm) \\ 
%\hline  &  & &  &\\  
\hline 144 & 15 & 15 & 17 & 17 \\ 
\hline 488 & 12 & 11 & 15 & 15 \\ 
\hline 3167 & 12 & 11 & 15 & 14 \\ 
\hline 7556 & 13 & 12 & 15 & 15 \\ 
\hline 
\end{tabular} 
\caption { Full width at half maximum of the proton beam as measured by arrays before and after a stack of approximately 40 300$\mu$m silicon wafers, at four instances during a run.}
 \label{tab:blood} 
\end{center}
\end{table}

At any particular point in the stack, the resolution on the proton beam profile depends on the diode density. Because the grid is constituted by discrete points, the resolution is given by $p/\sqrt{12}$, where $p$ is the pitch [10]. In our case $p= 3.8$ mm so the resolution is 1.1 mm. The diodes in these prototypes were soldered by hand which limited the pitch achievable. Future iterations of this device could use integrated diodes with higher density and consequently better resolution.

 \begin{figure}
 \begin{center}
\includegraphics[width=7in,height=3.6in]{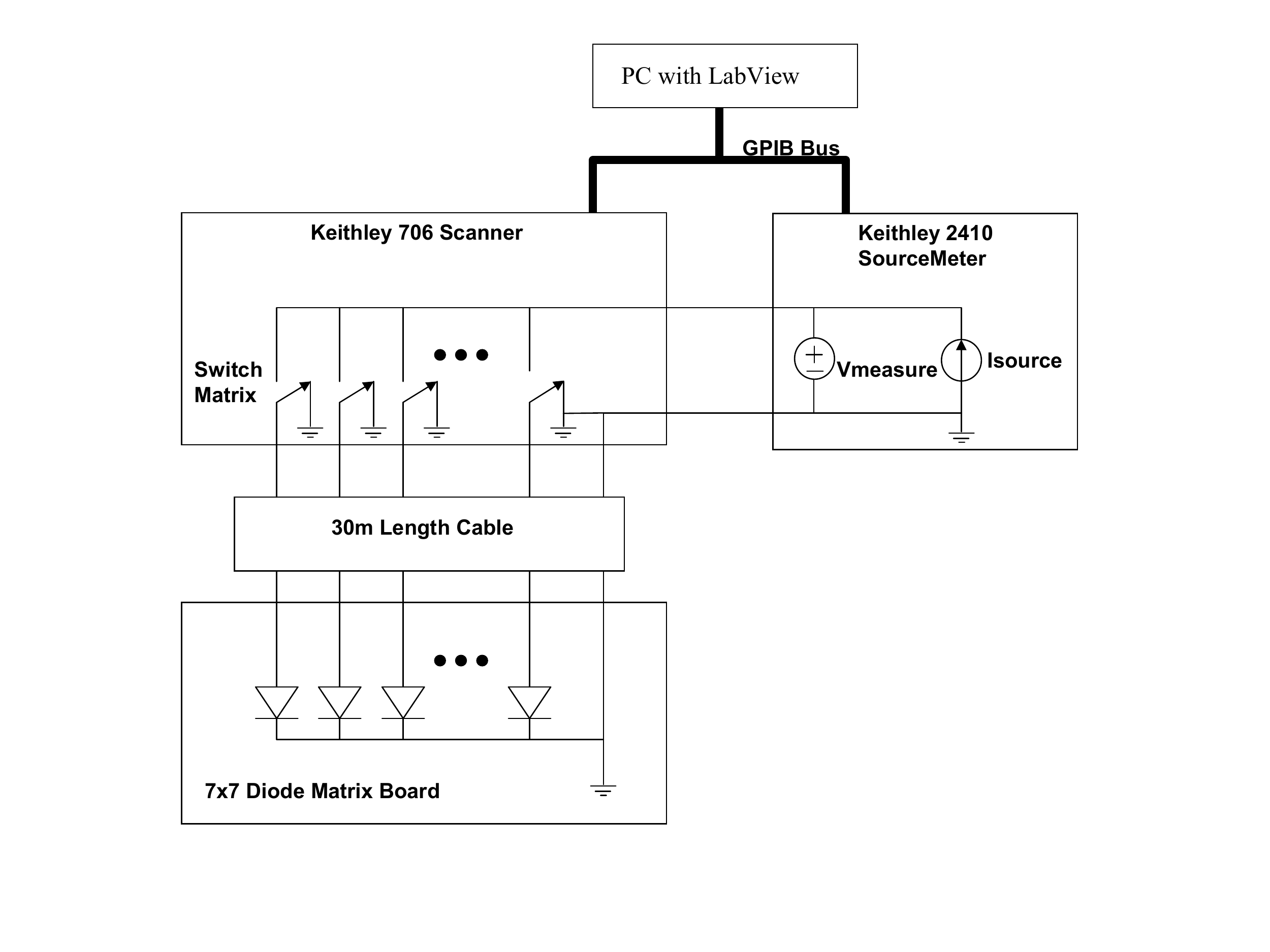}
\caption{ Electrical connections for the diode array readout. }
%\label{ }
\end{center}
\end{figure}
 \begin{figure}[t!]
 \begin{center}
\includegraphics[height=3.5in, width=4.5in]{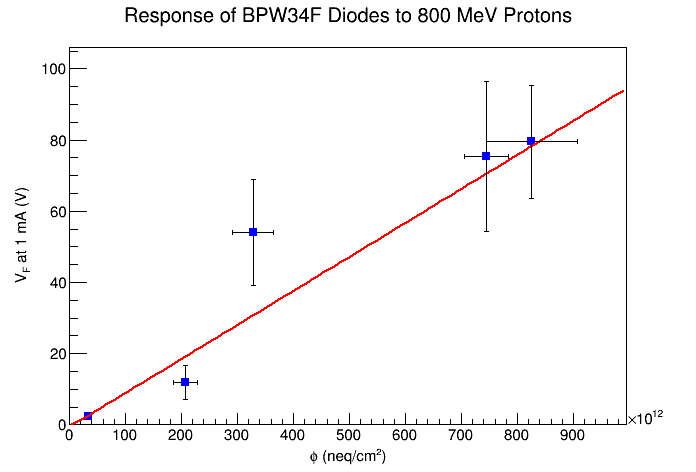}
\caption{ Calibration plot showing response of forward voltage of the diode array at 1 mA as a function of proton fluence. }
%\label{ }
\end{center}
\end{figure}

\section{Systematic Uncertainties}
The current pulse width contributes uncertainty on the diode voltage of about 5$\%$. This could be made arbitrarily smaller with use of a different sourcemeter. The cable from the diode array to the readout scanner contributes an uncertainty of under 9$\%$. This could be reduced by using a four wire measurement. The Keithley 2410 Sourcemeter measures the voltages to a precision of 0.015$\%$ + 50 mV. The uncertainty due to the temperature coefficient of the OSRAM BPW34F $p-i-n$ diodes is about 2.6$\%$. The total uncertainty in the measurement of the fluence using forward voltage is determined to be 11$\%$ by summing in quadrature the individual uncertainties. This is comparable to the uncertainty achievable from a 60 minute post-experimental count of an activated aluminum foil.

\begin{figure}[t!]
\begin{center}
\includegraphics[width=4.1in,height=3.4in]{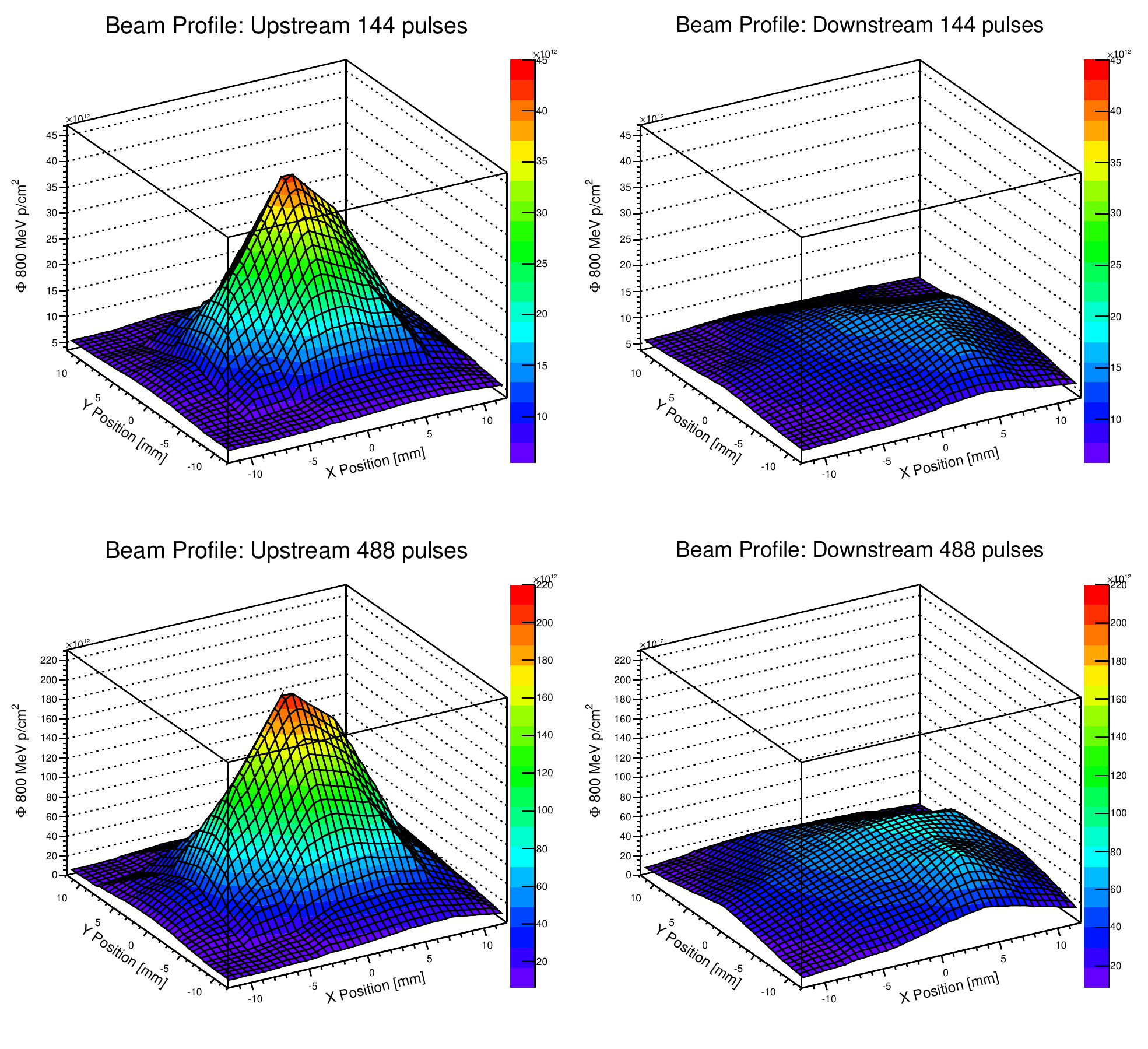}
\includegraphics[width=4.1in,height=3.4in]{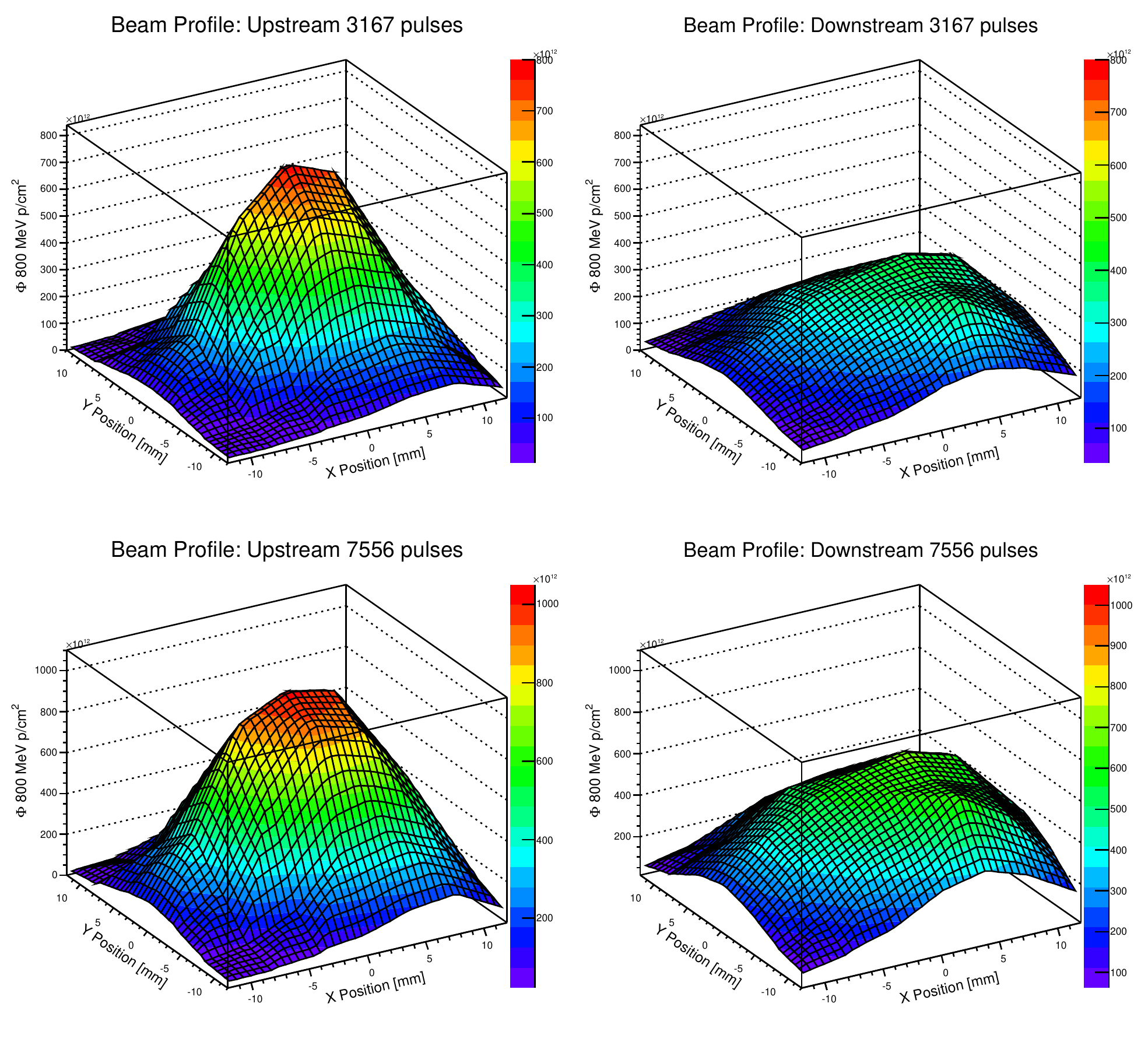}
\caption{ An example real time measurement of an evolving beam profile, made using a pair of diode arrays placed upstream (left column) and downstream (right column) of a stack of about 40 300-$\mu$m thick silicon sensors. In each subgraph, x and y indicate diode position. The vertical axis is fluence derived from voltage. The fluence received (in number of proton macro-pulses) increases downward. Note the vertical axes are different for each row. The graphs have been smoothed. }
\end{center}
\end{figure}

\section{Conclusions}
A method for rapid in-situ measurement of beam profile and fluence using a diode array system is described. The fluence calibration of the diode array has been confirmed using aluminum foils. A few seconds measurement of the fluence delivered during operation can be accomplished without stoppage of the beam to a precision (11$\%$) comparable to that from a $\sim$60 minute post-experimental count of an activated aluminum foil. Using this technique we have verified the deterioration of the beam profile as the beam traverses a stack of approximately 40 300-$\mu$m silicon sensors. The resolution on real-time measurement of the beam profile is limited only by the pitch at which the experimenter assembles the diodes.

% \begin{figure}
%\begin{center}
%\includegraphics[width=3in,height=3in]{./array1.jpg}
%\caption{ The front side of the diode array.}
%%\label{ }
%\end{center}
%\end{figure}

% \begin{figure}
%\begin{center}
%\includegraphics[width=3in,height=3in]{./foil.jpeg}
%\caption{ Aluminum foil matrix attached to the diode array.}
%%\label{ }
%\end{center}
%\end{figure}

%
%
% \begin{figure}
%\begin{center}
%\includegraphics[width=5in,height=4in]{./foil.eps}
%\includegraphics[width=5in,height=4in]{./pin.eps}
%\caption{ The upper plot shows the fluences measured by every element of the aluminum foil matrix. The lower plot shows the fluences measured by every element of the diode array to which the calibration foil was attached. These measurements were taken after 7556 macropulses of protons (about 3.7 x 10$^{15}$ protons) had been received.}
%%\label{ }
%\end{center}
%\end{figure}
%

\end{document}